\documentclass{article}

\usepackage[version=3]{mhchem}
\usepackage{siunitx}
\usepackage{graphicx}
\usepackage{tabularx}
\usepackage[a4paper,top=3cm,bottom=2cm,left=2.35cm,right=2.35cm,marginparwidth=1.75cm]{geometry}
\usepackage[sort&compress,numbers]{natbib}

\title{Enhancement of photoacoustic spectroscopy with sorption enrichment for ppt-level benzene detection}
\date{}
\author{}

\begin{document}
\twocolumn[
\maketitle
\begin{centering}
\vspace{-40pt}
Juho Karhu\textsuperscript{1,2,*} and Tuomas Hieta\textsuperscript{3}\\
\vspace{15pt}
\textsuperscript{1}Department of Chemistry, University of Helsinki, P.O. Box 55, 00014, Helsinki, Finland\\
\textsuperscript{2}Metrology Research Institute, Aalto University, Maarintie 8, 02150, Espoo, Finland\\
\textsuperscript{3}Gasera Ltd., Lemminkäisenkatu 59, FI-20520,
Turku, Finland\\
\textsuperscript{*}Corresponding author: juho.j.karhu@aalto.fi\\
\end{centering}
\vspace{5pt}
\hrule
\vspace{-2pt}
\begin{abstract}
\noindent A real time trace gas detector for benzene is demonstrated. The measurement system takes an advantage of modest enrichment through short adsorption periods to reach a ppt-level detection limit with a sampling cycle of \SI{90}{\second}, which includes sample adsorption, desorption and a spectroscopic measurement. Benzene is collected on Tenax TA sorbent for \SI{30}{\second} and then detected from the enriched samples with photoacoustic spectroscopy. A high sensitivity is achieved using cantilever-enhanced photoacoustic spectroscopy and a continuous-wave quantum cascade laser emitting at the wavelength \SI{14.8}{\micro\meter}, which corresponds to the absorption wavelength of the strongest benzene infrared band. We reach a detection limit of \SI{150}{ppt} of benzene, over one sampling cycle. Interference from humidity and other common petrochemicals is evaluated.

\vspace{6pt}
© 2022 Optica Publishing Group. One print or electronic copy may be made for personal use only. Systematic reproduction and distribution, duplication of any material in this paper for a fee or for commercial purposes, or modifications of the content of this paper are prohibited.
\end{abstract}
\vspace{6pt}
\hrule
\vspace{6pt}
]
\section{Introduction}
Benzene is a widely used chemical in industry, but it is known to be toxic and a carcinogen \cite{Smith10}. In addition to regulating industrial occupational exposure \cite{CAPLETON05}, benzene content in gasoline and its emission from traffic are concerns for public health \cite{Wallace96}. In addition to time spent in a car or in traffic, personal exposure is affected by sources such as tobacco smoke and residential garages \cite{WEISEL10,EDWARDS01,ILGEN01}. The current permissible occupation exposure limit is around \SI{1}{ppm} or below, depending on the regulatory body. The recommended exposure level by the US National Institute for Occupational Safety and Health is significantly lower at \SI{100}{ppb} and the Committee for Risk Assessment of the European Chemical Agency has recommended an occupation exposure limit of only \SI{50}{ppb}. For ambient exposure, the European commission has set a target limit value of \SI{1.5}{ppb}, evaluated as a yearly average. High sensitivity trace gas detectors for benzene are thus in high demand and for accurate monitoring a sub-ppb detection limit is needed for accurate quantification.

High sensitivities well below \SI{1}{ppb} can be reached with gas chromatography techniques \cite{LIAUD14}, but the measurements can be time consuming and the best performance often requires offline or offsite analysis with a separate sampling step \cite{ALLOUCH13}. Optical methods based on spectroscopic detection are an appealing alternative \cite{Popa21,Dumitras20}. Optical spectrometers are often more compact and more readily develop into field-capable instruments and autonomous onsite detectors for continuous monitoring \cite{Ueno01}. Some chemical sensors, such as metal oxide sensors, photoionization detectors and electrochemical sensors, are starting to reach high sensitivities for benzene \cite{Spinelle17}, but spectroscopic detectors offer better selectivity and long-term stability.

Various spectroscopic benzene detectors have achieved detection limits down to ppb-levels. However, many use the strong absorption band at the wavelength of approximately \SI{3}{\micro\meter} for detection \cite{Sur19,Cousin09,Mhanna21,hirschmann13}, which has significant overlap with many other organic compounds, most notably other simple aromatic combounds, e.g. toluene and xylenes. These are common interferents in benzene measurements and the requirement to properly account for their presence can degrade the performance of the detector substantially \cite{hirschmann13}. Another strong absorption band at \SI{14.8}{\micro\meter} has also been used for ppb-level benzene detection \cite{Young11,Shakfa21}, but previously a long preconcentration time of \SI{40}{\minute} was required for single digit ppb measurements \cite{Young11}.

In our previous publication, we were able to reach a sub-ppb detection limit without preconcentration, but only with a long averaging times of approximately \SI{20}{\min} \cite{Karhu20}. The high sensitivity was reached using cantilever-enhanced photoacoustic spectroscopy (CEPAS) for detection \cite{Kuusela07}. In CEPAS, the photoacoustic signal is recorded with an optically-read cantilever microphone, which leads to increased sensitivity compared to conventional diaphragm microphones. The technique has been used to reach sub-ppb detection limits for various trace gasses \cite{Peltola15,Tomberg18,Tomberg19,Fatima21}. In this article, we have significantly enhanced the detection limit of our previous spectrometer with sample enrichment using a Tenax TA sorbent for preconcentration. Sorbent based preconcentration couples well with CEPAS, because CEPAS allows highly sensitive measurements even with a small sample volume \cite{Karhu19,Tomberg20}. When limited gas volume is used for desorption, the spectroscopic measurement can be performed from a highly enriched sample volume. The CEPAS measurement is not done from a continuous flow, but instead the CEPAS cell is closed during a spectroscopic measurement. This makes the setup time efficient, because once the enriched sample has been transported into the CEPAS cell, the spectroscopic measurement can be performed in parallel to the collection of the subsequent sample into the sorbent. The measurement averaging time can be matched to the adsorption time to optimize the sensitivity. Using short adsorption times, we are able to reach a detection limit well below \SI{1}{ppb} with a \SI{90}{\second} sampling cycle. To our knowledge, this is the first time that CEPAS measurements of enriched samples have been reported.

\section{Experimental}
\subsection{Spectroscopic setup}
\label{sect:spectroscopy}
\begin{figure}[htbp]
\centering
\includegraphics[width=\linewidth]{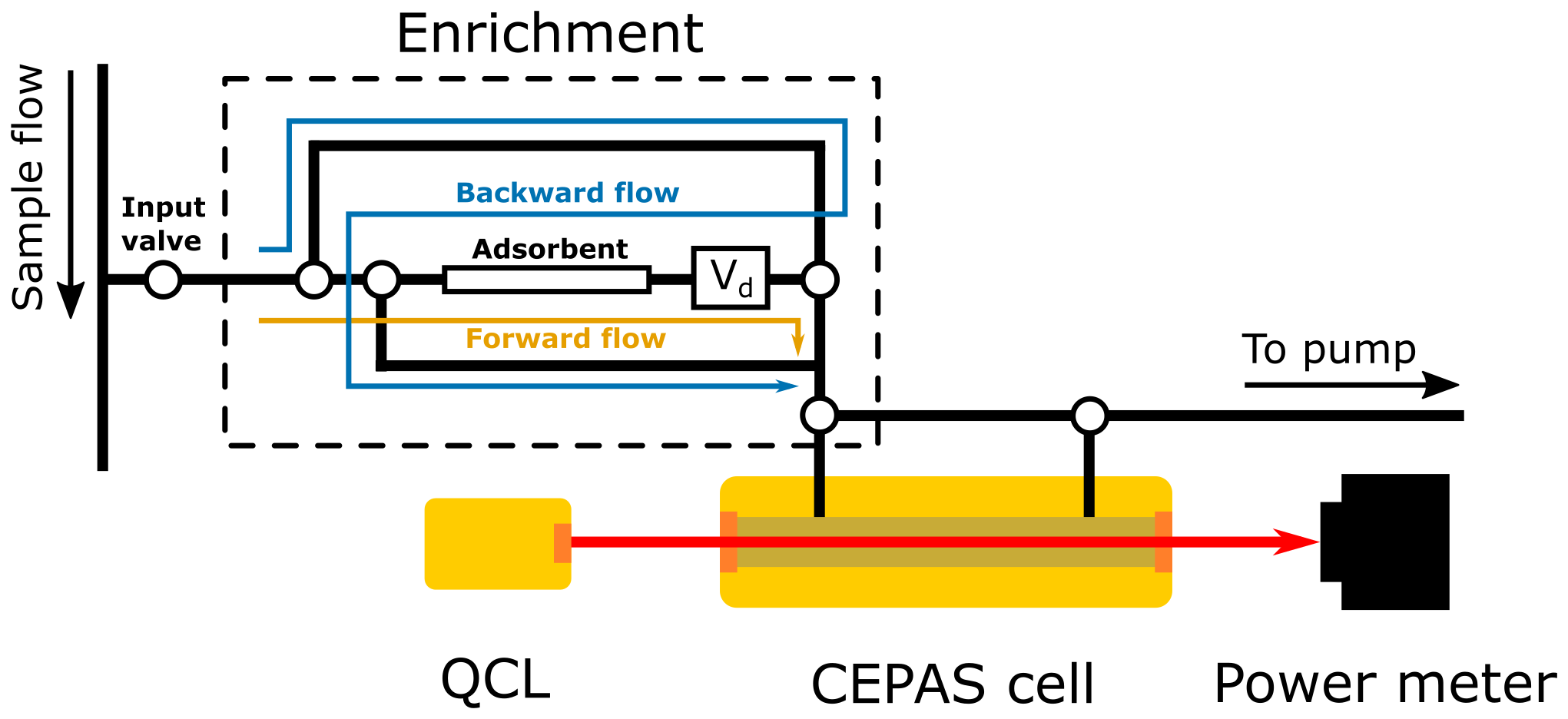}
\caption{Schematic picture of the measurement setup. Circles are voltage controlled solenoid valves. The valves are used to control the gas exchange and the direction of the flow through the sorbent. The forward flow is used during adsorption phase and the backward flow is used for desorption and flushing the sorbent. Symbol $V_d$ marks an extra \SI{10}{\milli\litre} volume used to assist in desorption.}
\label{fig:setup}
\end{figure}

\begin{figure}[htbp]
\centering
\includegraphics[width=\linewidth]{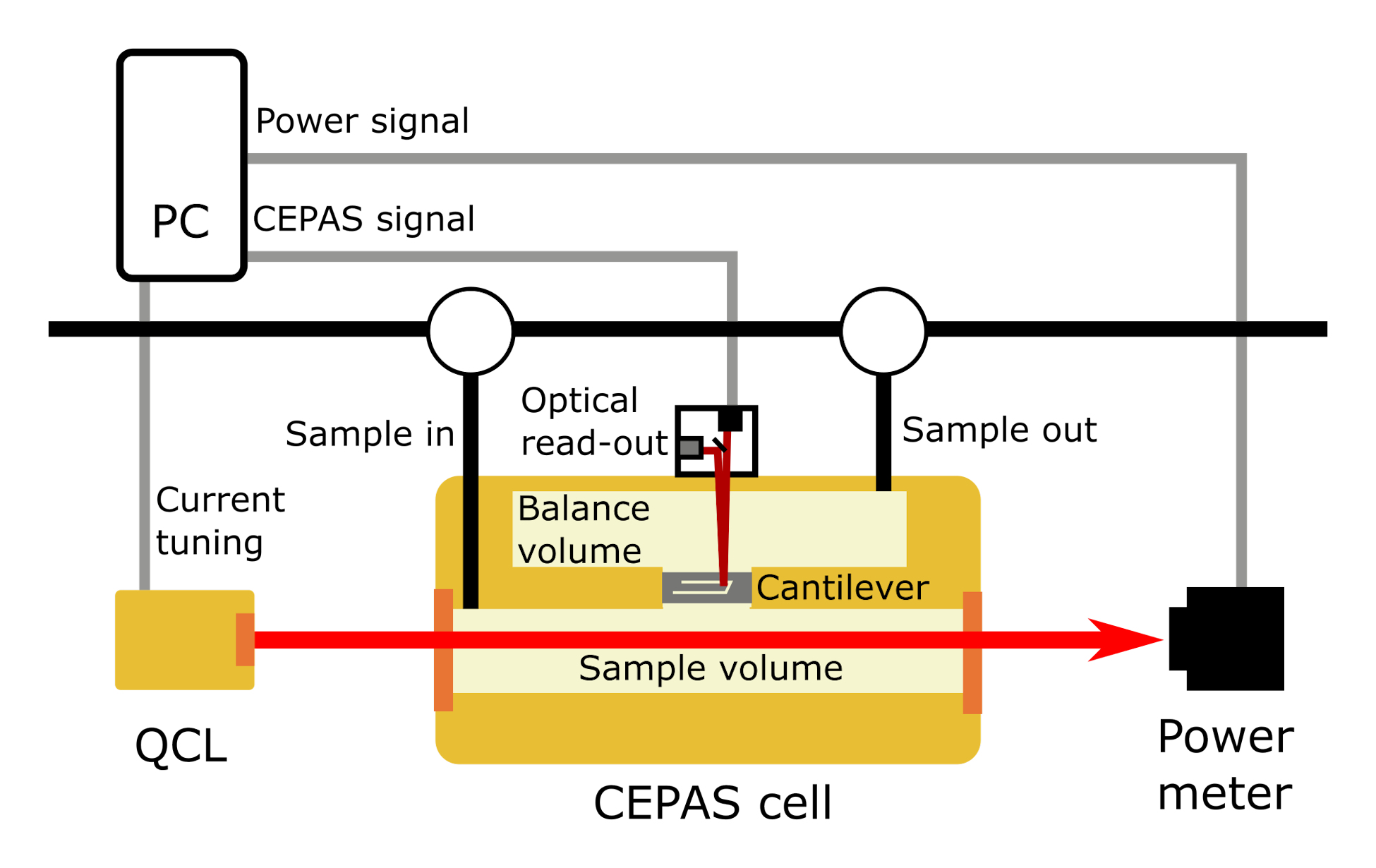}
\caption{Schematic picture of the spectroscopic part of the measurement setup. The spectroscopic detection is described in more detail in our previous article \cite{Karhu20}. The light source is a continuous-wave quantum cascade laser emitting at the wavelength \SI{14.8}{\micro\meter}. The laser passes through a sample cell equip with a cantilever microphone, which measures the generated photoacoustic signal. The cantilever movement is recorded optically with an interferometer. The microphone signal and the measured laser power are recorded with a computer running a LabVIEW measurement program. Laser wavelength tuning and modulation are controlled by the computer.}
\label{fig:setup_spectro}
\end{figure}

The measurement setup is depicted in Fig. \ref{fig:setup}. The spectroscopic part is similar to that described in our previous article \cite{Karhu20}, and a schematic picture of it is depicted in Fig. \ref{fig:setup_spectro}. In brief, it consists of a continuous-wave quantum cascade laser (QCL) and a photoacoustic cell containing an optically read cantilever microphone (PA201, Gasera). The laser is an InAs and AlSb based QCL with an emission wavelength of \SI{14.8}{\micro\meter} \cite{Karhu20,Baranov16,NguyenVan19}. The QCL used here was a custom build laser provided for this work by University of Montpellier, although now these lasers are commercially available (uniMir, mirSense). The operational temperature of the QCL is about 250 K, which is maintained with a water-cooled thermoelectric cooler. The output power is up to about 5mW. The laser beam passes through the CEPAS cell with anti-reflection coated \ce{ZnSe} windows. The volume of the cell is about \SI{8}{\milli\litre} and the cell temperature is stabilized to \SI{50}{\celsius}. The optical power after the photoacoustic cell is measured with a thermopile detector and the power reading is used to normalize the photoacoustic signal. For the spectral measurements, the QCL wavelength is tuned by its current. Wavelength modulation absorption spectroscopy (WMAS) is used for the detection \cite{KLUCZYNSKI01}, and the wavelength modulation is also applied through the QCL current tuning. At the start of a measurement cycle, first several points are measured at the peak of a \ce{CO2} line at \SI{674.075}{\per\centi\meter}, followed by a measurement of 23 spectral data points over the benzene absorption feature. The \ce{CO2} line is used to calibrate the wavelength axis, as well as to subtract the interference from the overlap of its tail with the benzene absorption band. The \ce{CO2} corrections are described in more detail in our previous article \cite{Karhu20}. To determine the benzene concentration in a sample, linear least squares method is used to fit the measured benzene spectrum with a reference spectrum, which has been recorded from a sample with a known benzene concentration.

\subsection{Enrichment stage}
The enrichment stage is added to the setup before the CEPAS cell as shown in Fig. \ref{fig:setup}. Directly after the sample input, the sample gas flows through a glass sorption tube containing about \SI{150}{\milli\gram} of Tenax TA (60/80 mesh, Alltech). Tenax TA does not adsorb benzene particularly strongly, but the relatively low breakthrough volume is not an issue here, because the adsorption time is short. Furthermore, this ensures that the desorption is very efficient even at a relatively low temperature, so that the duration of the total measurement cycle can be kept short. The flow direction through the adsorbent can be chosen with voltage controlled three-way valves. To control the sorbent temperature, a resistive heater is coiled around the sorbtion tube and two computer case fans are used for forced air cooling of the sorben tube. The temperature of the sorbent is monitored with a thermistor probe inserted inside the sorbent tube through a rubber seal.

\begin{figure*}[tb!]
\centering
\includegraphics[width=\linewidth]{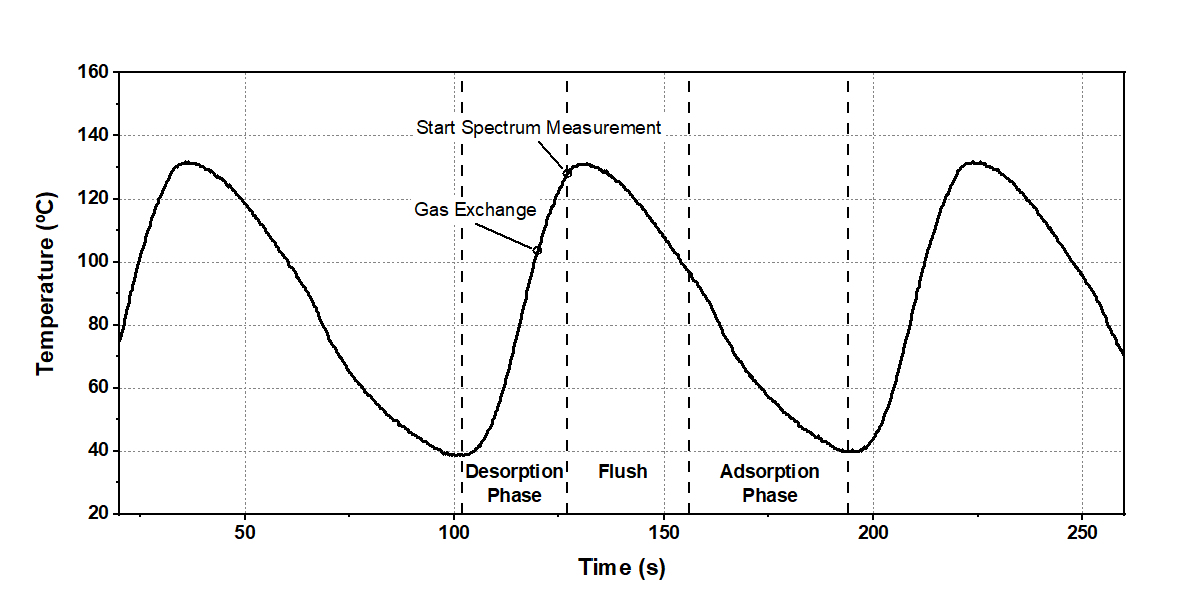}
\caption{Temperature of the sorbent over the measurement cycle. See text for detailed explanation of the cycle. The spectrum is measured in parallel with the flushing and collection of the subsequent sample, i.e. during Flush and Adsorption Phase. When the spectrum measurement is done, desorption phase is started. The sorbent is heated and when the temperature is high enough to ensure almost all benzene is desorbed, a new gas sample is moved into the photoacoustic cell (Gas Exhange).}
\label{fig:curve}
\end{figure*}

\begin{figure*}[htb]
\centering
\includegraphics[width=\linewidth]{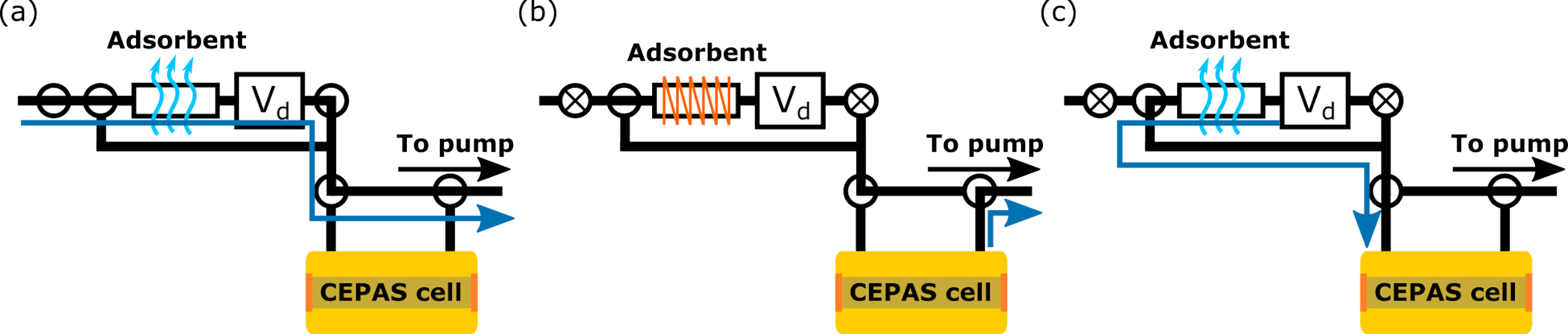}
\caption{Schematic picture of the gas exchange during desorption. The valves marked with X are closed two-way valves. (a) During adsorption phase, sample flows through the adsorbent and the sorbent tube is cooled with forced air cooling. The CEPAS cell is isolated from the gas line and the spectroscopic measurement is ongoing. (b) When the desorption phase starts, the sorbent tube is isolated and the heating starts. The CEPAS cell output valve is opened and the cell is evacuated. (c) When the temperature has risen sufficiently, the sorbent tube is connected to the CEPAS cell, the cell output valve is closed and the sorbent tube cooling is started. The enriched sample from the sorbent tube expands into the evacuated cell.}
\label{fig:desorb}
\end{figure*}

Duration of one measurement cycle is approximately \SI{90}{\second}. The cycle can be roughly divided into the desorption, flushing and adsorption phases. Figure \ref{fig:curve} demonstrates the measurement cycle and how the sorbent temperature evolves over time. It should be noted that the measurement is continuous and cyclic. Here, we have arbitrarily chosen to explain the sampling cycle from the point of view of the spectroscopic measurement, in the sense that we start from the desorption phase, where the sample exchange into the CEPAS cell takes place. The adsorption phase, which happens parallel to the CEPAS measurement, is explained last. In the desorption phase, an enriched sample is transferred from the sorbent tube into the CEPAS cell. During the transfer, there is no gas flow from the input through the sorbent. Instead, the gas from the sorbent tube is transferred passively into the cell by first evacuating the cell and then opening the valve between the evacuated cell and the sorbent tube. To ensure that there is enough gas volume to transfer the desorbing gases throughout the sorbent, there is an extra volume of about \SI{10}{\milli\litre} behind the sorbent tube (see $V_d$ in Fig. \ref{fig:setup}). Figure \ref{fig:desorb} depicts the steps of the gas exhange. It should be noted that Fig. \ref{fig:desorb} starts from the end of the adsorption phase (Fig. \ref{fig:desorb}a). In the beginning of the desportion phase, the valves before and after the sorbent tube are closed, and the heating of the sorbent tube is initiated (Fig. \ref{fig:desorb}b). While the sorbent is heating up, the CEPAS cell is evacuated to a pressure of \SI{150}{\milli\bar}. When the sorbent temperature is measured to be over \SI{100}{\celsius}, the heating is stopped and the forced air cooling of the sorbent tube is turned on. The flow direction is set into the backward flow direction (see Fig. \ref{fig:setup}) and the valve between the sorbent tube and the CEPAS cell is opened (Fig. \ref{fig:desorb}c). The input valve is still closed, but the gas from the extra volume $V_d$ expands through the sorbent towards the evacuated CEPAS cell, transporting the enriched sample gas into the cell. At this stage, the cell pressure rises to approximately \SI{440}{\milli\bar}. The valve between the cell and the sorbent tube is then closed, and the cell pressure is evacuated to the measurement pressure, which in our measurements is \SI{200}{\milli\bar}. The low pressure is required to reduce the spectral interference from the \ce{CO2} line. With the enriched sample inside the CEPAS cell, the spectrum measurement is initiated together with the flushing phase of the enrichment.

Sorbent is flushed into backwards direction (see Fig. \ref{fig:setup}) for approximately \SI{30}{\second} after the desorption phase, when the sorbent temperature is still high. Here we used sample gas for flushing, since very little benzene is retained in the sorbent when the temperature is well over its boiling point. In principle, better accuracy could be expected if, for example, clean nitrogen or another input equipped with an active carbon filter was used for the flushing, but in most practical cases we expect the improvement to be marginal. After the flushing phase, the adsorption phase is initiated by switching the flow direction to the forward flow (see Fig. \ref{fig:setup}). It should be noted that the sorbent temperature is changing during the adsorption phase in our measurements. In the beginning, the temperature is still so high that a part of the adsorption time is effectively wasted, since the breakthrough volume is small and very little benzene is absorbed. According to our measurements, we estimate that once the sorbent temperature drops below \SI{80}{\celsius}, practically all of the benzene is retained. The effective adsorption time where the sorbent temperature is below \SI{80}{\celsius} in our measurements is approximately \SI{30}{\second}. During the flushing and adsorption phases, which take approximately \SI{60}{\second}, the sample spectrum is measured from an enriched sample inside the CEPAS cell and the benzene concentration is calculated as described in the previous section. After the spectrum is measured, the gas flow is stopped and the desorption phase is initiated again, starting a new measurement cycle.

The measurement system is constantly monitoring the sorbent temperature and the flow rate at the sample input, since these have an impact on how much benzene is retained in the sorbent. The sample flow rate into the measurement system during the adsorption phase is approximately \SI{0.5}{\litre\per\minute}. In the laboratory measurements reported here, we didn't observe much variation in these parameters, and the repeatability was good without any corrections. In field measurements however, for example if the ambient temperature varies significantly, it may be necessary to calibrate the sorbent enhancement with factors tied to the sorbent temperature and the sample mass flow rate.

\section{Results and Discussion}
\begin{figure}[htb!]
\centering
\includegraphics[width=\linewidth]{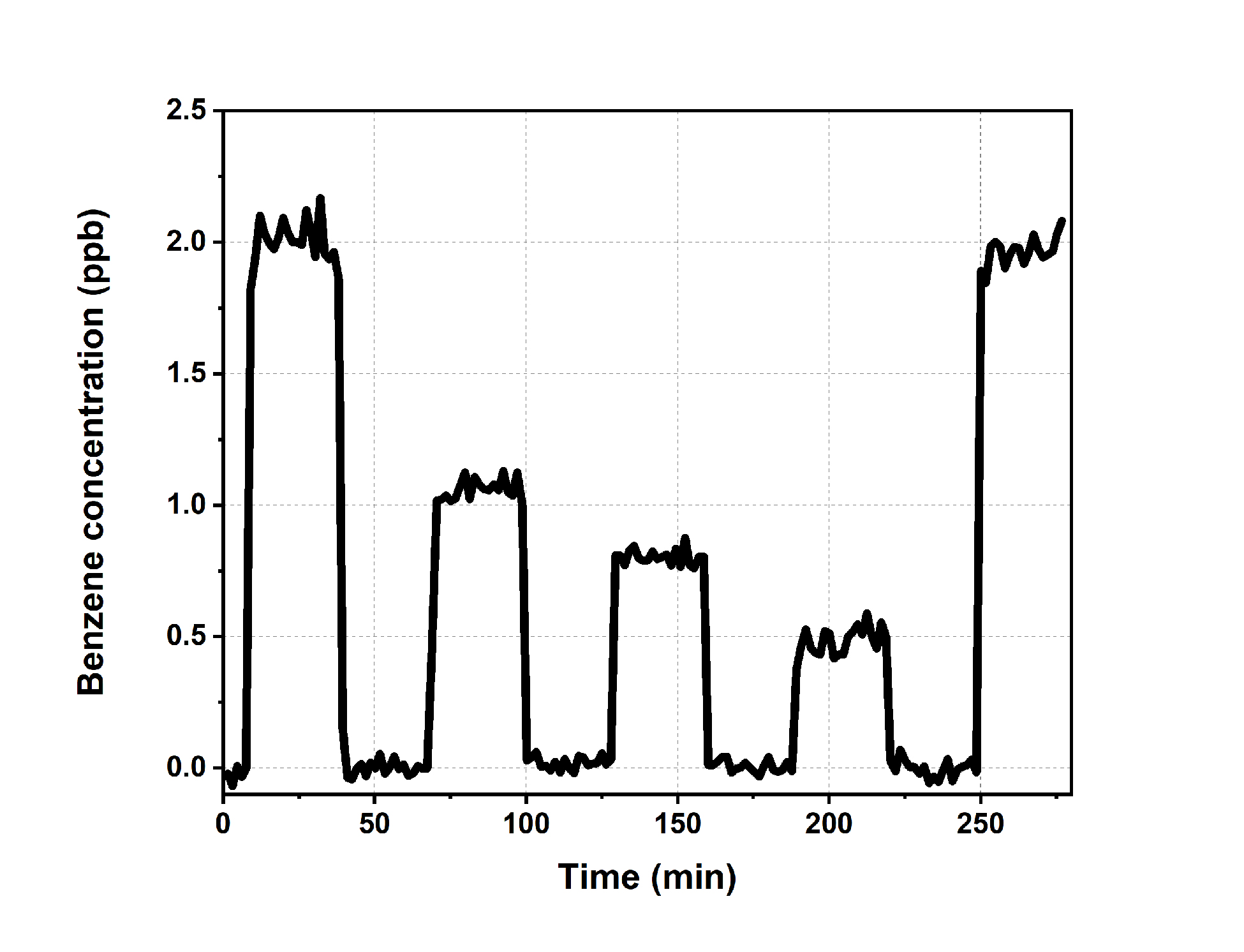}
\caption{Detected benzene concentration from a measurement set with varying benzene levels. The benzene concentration in the sample for the four steps were \num{2}, \num{1}, \num{0.8} and \SI{0.5}{ppb}. At the end, the concentration was again set to \SI{2}{ppb} to demonstrate the repeatability over a measurement period of \SI{4.5}{\hour}. In between the steps, the benzene concentration was set to zero.}
\label{fig:steps}
\end{figure}

To generate benzene samples with known concentrations, cylinder gas containing 1 ppm of benzene in nitrogen balance was diluted with compressed air. The compressed air was first sent through an active carbon column filter to ensure it is free of benzene. The mixing ratio was controlled with mass flow controllers. A flow of \SI{1}{\litre\per\minute} of the benzene mixture was flowing pass the sample input of the measurement setup and the flow was sampled by the instrument in bypass mode. Figure \ref{fig:steps} shows the response of the measurement system to varying benzene concentrations. Each point corresponds to one \SI{90}{\second} cycle. The benzene concentration in the sample flow was kept constant for a duration of \SI{30}{\minute} and then changed to a new level. Between each \SI{30}{\minute} constant concentration measurement set, a blank measurement with no benzene was performed from a separate gas flow containing only the filtered compressed air. This allowed us to verify that the zero level was retained over the measurement set. During the blank measurement, the benzene concentration in the mixed flow had time to stabilize to the new level. This process was repeated several times over a \SI{4.5}{\hour} measurement set.

\begin{figure}[htb!]
\centering
\includegraphics[width=\linewidth]{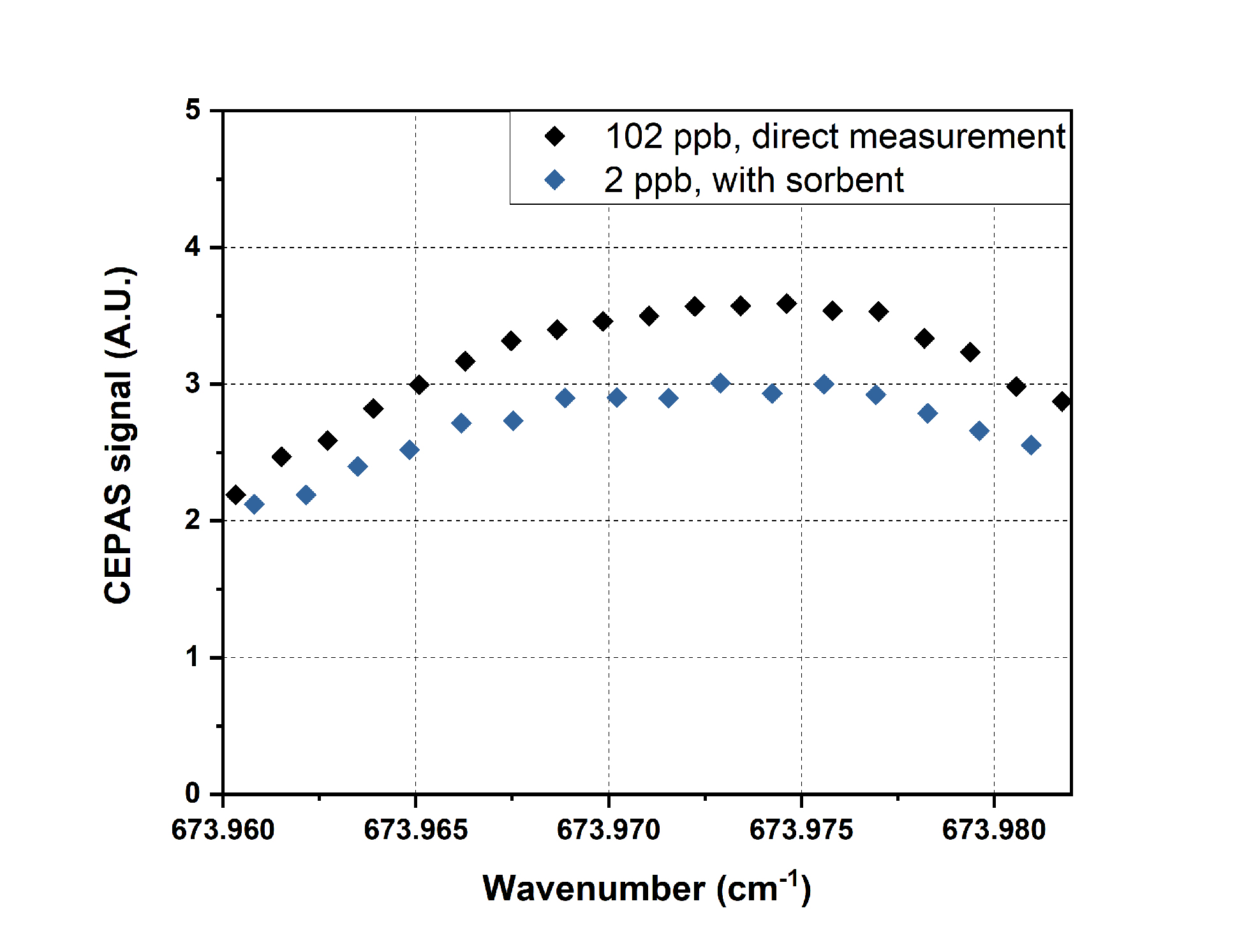}
\caption{Comparison of the benzene photoacoustic spectrum with and without the adsorption enrichment. The \ce{CO2} contribution has been subtracted from the measured spectra as described in the text. Both spectra were averaged over 16 spectrum measurements. The spectrum measured with enrichment from a sample flow containing only \SI{2}{ppb} of benzene (blue) is of the same magnitude as a spectrum measured without the enrichment from a sample flow containing \SI{102}{ppb} of benzene (black).}
\label{fig:ads_vs_direct}
\end{figure}

By comparing the measured benzene spectra to one measured with the same spectrometer without using the enrichment, we can determine that the benzene concentration in the enriched samples inside the CEPAS cell was approximately 57 times larger than the benzene concentration in the sample flow (see Fig. \ref{fig:ads_vs_direct}). That is, the enrichment offered an enhancement factor of 57. The flow rate into the measurement device was about \SI{0.5}{l/min}. We estimated before that the effective adsorption period was approximately \SI{30}{\second}. Therefore, we estimate that benzene contained in about \SI{0.25}{\litre} volume of sample was captured in one adsorption period. If we further assume that all of this was transported into the CEPAS cell by the gas expanding through the sorbent tube during the desorption step, we can calculate a maximum for the predicted enhancement factor. This assumption would mean that benzene is only moving towards the CEPAS cell with the flow from the volume $V_d$ and that all of it escapes the adsorbent, which is clearly an overestimation. The final pressure inside the CEPAS cell was about \SI{440}{\milli\bar} and the cell volume is about \SI{8}{\milli\litre}. Ignoring the temperature difference of about \SI{20}{\kelvin}, the benzene contained in a gas volume of \SI{0.25}{\litre} was effectively measured from a gas volume of \SI{3.52}{\milli\litre}, giving an enhancement factor of 71. This is an estimate for the maximum attainable enhancement factor and it is in reasonable agreement with our experimental value, which was 57 as described above. We believe the reduction from the maximum value is mostly due to the desorption volume not transporting all of the adsorbed benzene out of the sorbent.

\begin{figure}[htb!]
\centering
\includegraphics[width=\linewidth]{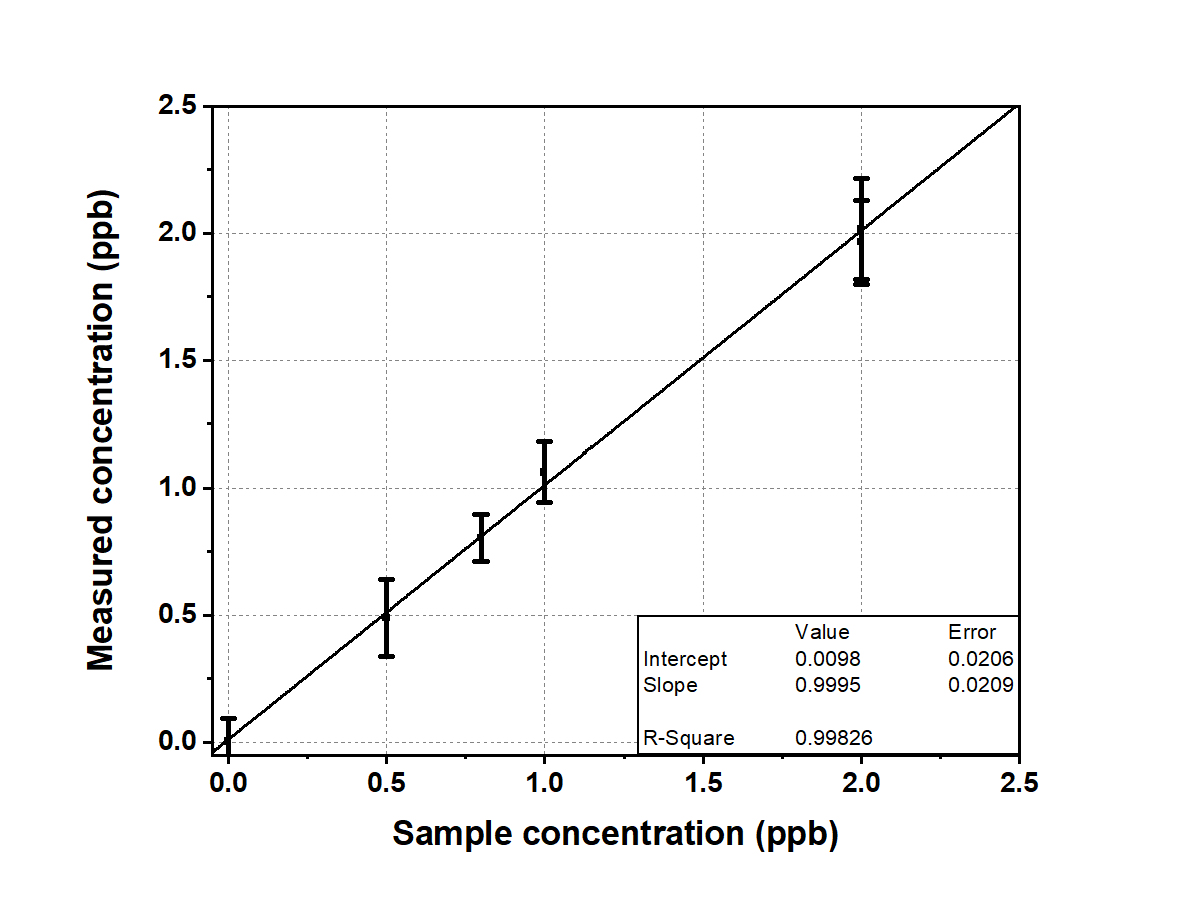}
\caption{Concentration response, as a function of the known sample concentrations. Each point was measured as an average over \SI{30}{\min}. The error bars are three standard deviations of the measurement sets.}
\label{fig:calib}
\end{figure}

The average benzene concentrations for the concentration steps in Fig. \ref{fig:steps} were used to construct a calibration curve shown in Fig. \ref{fig:calib}. The error bars equal three times the standard deviation ($3\sigma$) of the data at each step. The calibration curve demonstrates very good linearity over the measured concentration range. As an estimate for a detection limit, three times the standard deviation for the lowest measured concentration step (\SI{0.5}{ppb}) was \SI{150}{ppt}. The largest standard deviation within one concentration step was for the first level, where one standard deviation ($1\sigma$) corresponded to a benzene concentration of \SI{66}{ppt}. For the blank measurements, standard deviation was \SI{30}{ppt} or less within one individual \SI{30}{\minute} measurement set. Between the different \SI{30}{\minute} blank measurement sets, the largest difference between two average values was \SI{27}{ppb}, suggesting there is slight drifting in the background level over the measurement period of \SI{4.5}{\hour}. The parameters representative of the measurement system are collected in Table \ref{tab:lod}.

\begin{table}[h!]
\caption{Performance parameters of the adsorption enhanced photoacoustic benzene sensor}
\centering
\begin{tabularx}{\linewidth}{| >{\raggedright\arraybackslash}X |>{\centering\arraybackslash}X|}
\hline
 \centering Parameter & Value \\
\hline
LOD\textsuperscript{a} & 150 ppt \\ 
Sampling time & 90 s\\
Drift\textsuperscript{b} & 27 ppt over 4 h  \\
\hline
\end{tabularx}
\\
\footnotesize
\textsuperscript{a} 3$\sigma$ standard deviation for measurement of 0.5 ppb of benzene.\\
\textsuperscript{b} Largest difference between 30 min blank measurements.\\
\label{tab:lod}
\end{table}

A longer adsorption time can be used for higher enhancement at a cost of response time. However, with Tenax TA as the absorbent, the breakthrough volume of benzene is limited, so a different choice of adsorbent might be necessary if long adsorption times are desired. For online measurements with relatively fast response, Tenax TA provided a good compromise, since it allows fast desorption at relatively low temperature for benzene. In our laboratory measurements, the repeatability of the sampling cycle was good, despite part of the adsorption taking place in varying temperature. In a more demanding measurement environment, the reliability of the sampling cycle could be improved, again at a cost of response time, by first cooling the sorbent to a set temperature before starting the adsorption phase. We also tested somewhat shorter sampling times. By reducing the duration of the adsorption phase so that the total cycle period was \SI{1}{\minute}, the enhancement dropped to a factor of 23. With the shorter cycle, the effective adsorption time, i.e. the part of the adsorption phase where the sorbent temperature was below \SI{80}{\celsius}, was approximately \SI{12}{\second}.

\subsection{Interference measurements}
The interference from \ce{CO2} is treated by a correction applied to the spectral measurement, as described in section \ref{sect:spectroscopy}, and it was shown to be effective in our previous article \cite{Karhu20}. In the interference measurements performed here, the shorter sampling cycle of \SI{1}{\minute} was used. We first evaluated the interference from humidity. The benzene concentration in the sample flow was set to a constant level of \SI{3}{ppb}. The air supply used to dilute the benzene was first sent through a water evaporator to control the water concentration in the sample flow. Benzene response was measured with four water concentration levels ranging from 0 to \SI{2.5}{\percent} in mole fraction. The water concentration in the sample flow was verified with a humidity meter (ChipCap 2-SIP, Telaire). Each humidity level was measured for \SI{20}{\minute}. A small decrease in the measured benzene concentration as a function of increasing humidity was observed. The benzene concentration decreased by approximately \SI{0.14}{ppb} when the water concentration increased by \SI{1}{\percent} in mole fraction.

The interference from toluene, ethylbenzene and the xylene isomers was also evaluated. The samples were generated by passing a flow of compressed air, which was first purified with an active carbon filter, over the air space of liquid sample of each of the interferents. The resulting flow was assumed to contain a concentration in the order of the vapor pressure of the interferent. The flow was then diluted to a ratio of 1/1000 with compressed air. The diluted flow was sampled with the benzene measurement setup for \SI{20}{\minute} for each interferent. The interference was evaluated as the apparent benzene concentration reported by the instrument. The presence of the interferents in the diluted flow was verified with a proton transfer reaction time-of-flight mass spectrometer (PTR-TOF 1000, Ionicon). However, the instrument was not calibrated to the specific compounds under analysis and hence we could not verify the concentrations. The results of the interference measurements should be taken as order of magnitude estimates.

For all xylene isomers, any interference was below quantification and the equivalent benzene concentration remained zero to within the standard deviation of the measurement. For toluene, the average equivalent benzene concentration over the \SI{20}{\minute} measurement was \SI{0.166}{ppb}. The concentration of the main ion (\ce{(C7H8)H+}) from the mass spectrum was approximately \SI{3.8}{ppm}. When using this as the estimate for the toluene concentration, the relative interference signal is approximately \num{4e-5}. That is to say, \SI{1}{ppm} of toluene in the sample flow would result in an erroneous reading of \SI{40}{ppt} of benzene. For ethylbenzene, a higher interference signal corresponding to \SI{16.8}{ppb} benzene concentration was measured. However, unlike with toluene, the spectrum measured from the ethylbenzene sample matched the benzene reference spectrum very well and we believe that the interference was due to benzene impurity in the ethylbenzene sample (\SI{99.8}{\percent} pure, ACROS organics), since benzene is one of the most likely impurities. The peak corresponding to benzene was detected in the mass spectrum as well, but we cannot say with certainty what is the contribution of possible fragmentation and actual benzene impurity to the mass peak. A more pure ethylbenzene sample was not available to us at this time. In the ethylbenzene measurement, the concentration of the main ion (\ce{(C8H10)H+}) in the mass spectrum was \SI{750}{ppb}, so if we did assume that the interference was actually from ethylbenzene, the relative interference would be approximately \num{0.02}.

\section{Conclusion}
We have demonstrated a highly sensitive trace gas sensor for real-time detection of benzene. With a \SI{90}{\second} sampling cycle, we reach a detection limit of \SI{150}{ppt} (3$\sigma$). Common interferents were found to have little effect on the benzene measurements, but for accurate measurements of low benzene concentrations in the ambient air, it may be necessary to measure humidity of the sample flow and apply a correction to the measured concentration. Compact size and simple design of the measurement setup are ideal for development towards a portable instrument for on-site measurements. The small form factor and a simple measurement principle based on fitting of reference absorption spectra, support the potential of its use also in autonomous monitoring applications. Our results also show great compatibility between CEPAS and enhancement through enrichment. The CEPAS cell requires a sample volume of less than \SI{10}{\milli\litre} and as long as the desorption can be performed efficiently with equally small gas volume, the spectral measurement can be performed from a highly enriched sample.

\section*{Acknowledgements}

The authors would like to thank Matti Jussila and Kari Hartonen for useful discussion and assistance in assembling the sorbent tube, and Markus Metsälä for assistance with the mass spectrometer measurements. The authors would like to thank Roland Teissier, Alexei Baranov and Hadrien Phillip for providing the QCL used in this work. This work received funding from Jenny and Antti Wihuri foundation. This work was supported by the Academy of Finland Flagship Programme, Photonics Research and Innovation (PREIN), decision number: 320167.

\bibliographystyle{achemso}
\bibliography{bibliography}

\end{document}